\documentclass[twocolumn, showpacs, preprintnumbers,amsmath,amssymb,balancelastpage]{revtex4}
\pdfoutput=1

\usepackage{putexPRL}
\usepackage{graphicx}
\usepackage{dsfont}
\usepackage{amsmath}
\usepackage{array}
\usepackage{epstopdf}
\usepackage{enumerate}
\usepackage{youngtab}
\usepackage{tensor}
\usepackage{braket}
\usepackage[normalem]{ulem}
\usepackage{slashed}
\usepackage[aligntableaux=center]{ytableau}
\usepackage[utf8]{inputenc}
\usepackage[
      colorlinks=true,
      linkcolor=blue,
      urlcolor=blue,
      filecolor=black,
      citecolor=red,
      ]{hyperref}
\usepackage{braket}
\usepackage{mathtools}
\usepackage{rotating}
\usepackage{tabularx}
\newcolumntype{Y}{>{\centering\arraybackslash}X}
\newcolumntype{C}[1]{>{\centering\arraybackslash}p{#1}}
\usepackage{todonotes}
\usepackage{xcolor}
\definecolor{LightCyan}{rgb}{0.7,1,1}
\definecolor{Gray}{gray}{0.9}
\usepackage{xspace}

\newcommand{\abs}[1]{\left\lvert #1 \right\rvert}

\newcommand {\be} {\begin {equation}}
\newcommand {\ee} {\end {equation}}

\newcommand {\bes} {\begin {equation*}}
\newcommand {\ees} {\end {equation*}}

\newcommand{\es}[2] {\begin{equation} \label{#1} \begin{split} #2 \end{split} \end{equation}}

\newcommand{\Z}{\mathbb{Z}}

\newcommand{\beq}{\begin{equation}}
\newcommand{\eeq}{\end{equation}}

\def\ie{\begin{equation}\begin{aligned}}
\def\fe{\end{aligned}\end{equation}}

\def\<{\langle}
\def\>{\rangle}

\def\beg{\begin{equation}\begin{gathered}}
\def\eeg{\end{gathered}\end{equation}}

\def\bea{\begin{equation}\begin{aligned}}
\def\eea{\end{aligned}\end{equation}}

\begin{document} 

\preprint{PUPT-2634}

\title{Discrete Chiral Symmetry and Mass Shift\\ in Lattice Hamiltonian Approach to Schwinger Model}

\author{{Ross Dempsey,$^{1}$ Igor R.~Klebanov,$^{1,2}$ Silviu S.~Pufu,$^{1}$ and  Bernardo Zan}$^{1}$}         

\affiliation{$^{1}$Joseph Henry Laboratories, Princeton University, Princeton, NJ 08544, USA}
\affiliation{$^{2}$Institute for Advanced Study, Princeton, NJ 08540, USA}

\begin{abstract}
We revisit the lattice formulation of the Schwinger model using the Kogut-Susskind Hamiltonian approach with staggered fermions. This model, introduced by Banks et al., contains the mass term $m_{\rm lat} \sum_{n} (-1)^{n} \chi^\dagger_n \chi_n$, and setting it to zero is often assumed to provide the lattice regularization of the massless Schwinger model. We instead argue that the relation between the lattice and continuum mass parameters should be taken as $m_{\rm lat}=m- \frac 18 e^2 a$. The model with $m=0$ is shown to possess a discrete chiral symmetry that is generated by the unit lattice translation accompanied by the shift of the $\theta$-angle by $\pi$. While the mass shift vanishes as the lattice spacing $a$ approaches zero, we find that including this shift greatly improves the rate of convergence to the continuum limit. We demonstrate the faster convergence using both numerical diagonalizations of finite lattice systems, as well as extrapolations of the lattice strong coupling expansions.

\end{abstract}


\maketitle
\nopagebreak

\section{Introduction}

The $1+1$ dimensional model of Quantum Electrodynamics coupled to a fermion of charge $e$ and mass $m$, also known as the Schwinger model \cite{Schwinger:1962tp}, is a classic example of Quantum Field Theory. It is exactly solvable in some limits, and it is a very useful theoretical laboratory for various important phenomena including 
the confinement of charge. For $m=0$ the theory is exactly solvable \cite{Lowenstein:1971fc,Casher:1974vf, Coleman:1975pw}, reducing to the non-interacting Schwinger boson of mass $M_S\equiv e/\sqrt \pi$. The $U(1)$ chiral symmetry of the massless action is broken by the Schwinger anomaly. While the massive Schwinger model is not solvable exactly, a lot is known about it from the small mass and large mass expansions. In addition to containing the obvious dimensionless parameter $m/e$, the massive model may be generalized to include the $\theta$ angle related to introduction of the background electric field \cite{Coleman:1975pw}. This parameter, which has periodicity $2 \pi$, is somewhat analogous to the $\theta$ angle of the $3+1$-dimensional gauge theory. 

Since the massive Schwinger model is not exactly solvable, it is useful to introduce its non-perturbative lattice regularization. In this paper we will use the
Kogut-Susskind Hamiltonian approach with staggered fermions~\cite{Kogut_Susskind}, which in general resembles models relevant to condensed matter physics. The lattice Hamiltonian approach was originally applied to the Schwinger model by Banks, Susskind, and Kogut \cite{Banks:1975gq}, who initiated its strong coupling expansion. A major simplification in $1+1$ dimensions is that all the local gauge field degrees of freedom are eliminated through the Gauss Law constaints. This approach was further developed in many papers including \cite{Carroll:1975gb,Hamer:1982mx,Hamer:1997dx,Berruto:1997jv}. 

During the more recent period, the Lattice Hamiltonian approach to the Schwinger model has been attracting renewed attention. 
On the theoretical side, it is an excellent testing ground for advanced numerical techniques using Matrix Product States (MPS) and Density Matrix Renormalization Group (DMRG).
Using such approaches, various observables were computed with high accuracy \cite{Byrnes:2002gj,Byrnes:2002nv,Banuls:2013jaa,Buyens:2013yza,Buyens:2015tea,Banuls:2016lkq,Ercolessi:2017jbi,Funcke:2019zna,Magnifico:2019kyj,Honda:2021aum}.
Furthermore, there are exciting efforts to implement quantum simulations of the Schwinger model using various experimental setups 
\cite{Banerjee:2012pg,Martinez:2016yna,2017Natur.551..579B,Kokail:2018eiw,Klco:2018kyo,Surace:2019dtp,Yang:2020yer,Zhou:2021kdl} (for a review, see \cite{Banuls:2019bmf}).

In this paper we revisit the lattice Hamiltonian for the Schwinger model:
\es{eq:HLattice}{
	H &= \frac{e^2 a}{2} \sum_{n=0}^{N-1} \left( L_n + \frac{\theta}{2 \pi} \right) ^2 + m_{\rm lat} \sum_{n=0}^{N-1} (-1)^{n} \chi^\dag_n \chi_n  \\
	&- \frac{i}{2a} \sum_{n=0}^{N-1} \left[ \chi_n^\dag U_n \chi_{n+1}- \chi_{n+1}^\dagger U_n^\dagger \chi_n \right]  \,,
}
with $N$ even.  Here, $\chi_n$ and $\chi_n^\dagger$ are fermion annihilation and creation operators at site $n$, while $U_n=\exp(i \phi_n)$ and $U_n^\dagger = U_n^{-1}$ are unitary operators living on links between site $n$ and $n+1$. The electric field variables, also living on links, are $L_n=-i\frac{\partial}{\partial \phi_n}$; in the 
Hamiltonian (\ref{eq:HLattice}) they take integer values.
The parameters $e$, $a$, $\theta$, and $m_\text{lat}$ are the electric charge, lattice spacing, $\theta$-angle, and lattice mass parameter, respectively.   The Hamiltonian \eqref{eq:HLattice}  must be supplemented by the Gauss law constraints coming from gauge invariance, which are usually taken to be \cite{Hamer:1997dx}
\es{Gauss}{
	L_{n}-L_{n-1} = Q_n \,, \qquad Q_n \equiv \chi_n^\dag \chi_n -\frac{1-(-1)^n}{2}  \, .
}
For more details, see Section~\ref{sec:lattice} and Appendix~\ref{sec:HamiltonianAppendix}.

Our main new result is that the lattice mass parameter $m_{\rm lat}$, which originally \cite{Banks:1975gq,Hamer:1997dx} was not distinguished from the continuum mass $m$, should instead be identified as 
\es{mLat}{
	m_{\rm lat}=m- \frac 18 e^2 a \,, 
} 
where $a$ is the lattice spacing. While this mass shift vanishes in the continuum limit $a\rightarrow 0$, we will show that it greatly improves the rate of convergence to the continuum limit. The underlying reason is that the improved lattice definition of the $m=0$ theory respects a discrete remnant of the chiral symmetry. In the staggered formulation, where each two-component fermion is defined on a pair of adjacent sides, this discrete chiral symmetry is generated by the lattice translation by a single site 
\cite{Banks:1975gq,Susskind:1976jm}, and we find that it maps $\theta \to \theta + \pi$.
This is distinct from the spatial translation symmetry whose generator is translation by two sites; it leaves all parameters unchanged. We will work on a lattice of $N$ sites 
($N$ is an even integer)
with periodic boundary conditions, so that these lattice translations are not broken by the boundary conditions.  
Let us note that the lattice definition of the massless Schwinger model that is symmetric under the translation by one site has already appeared in a somewhat different guise 
\cite{Diamantini:1992ei,Berruto:1997jv}. The modification used there involved a different definition of the fermionic charges $Q_n$ than that in (\ref{Gauss}). In fact, as we demonstrate in Appendix~\ref{sec:HamiltonianAppendix}, the modification in \cite{Diamantini:1992ei,Berruto:1997jv} is equivalent to the original formulation of the model given in
(\ref{eq:HLattice}) and (\ref{Gauss}), but with the shifted mass. This allows us to make contact with the subsequent work on the lattice Schwinger model, which mostly used the original definition of \cite{Banks:1975gq,Hamer:1997dx} rather than the modified definition of \cite{Diamantini:1992ei,Berruto:1997jv}.

In section \ref{sec:lattice} we show that the strong coupling expansions, which can be carried out directly in the infinite volume limit \cite{Banks:1975gq,Carroll:1975gb,Hamer:1997dx}, exhibit excellent convergence to the known continuum results after the mass shift (\ref{mLat}) is taken into account.
In section \ref{sec:numerics} we also show that this improved identification of parameters leads to a good convergence to the known continuum results after carrying out exact diagonalizations with moderate values of $N$.
Besides using the chirally symmetric lattice definition of the massless theory, we deform the Hamiltonian via turning on the mass and study the dependence of various quantities on $m$ and $\theta$. 

In the model described above, the fermions have unit charge and the parameter $\theta$, with periodicity $2 \pi$, labels distinct ``theories,'' so the translation by one site is not strictly speaking a symmetry.  However, for any positive integer $q$, writing $U_n = (U_n')^q$, $L_n = L_n' / q$, $e = e' q$, $\theta = \theta' / q$, $m_\text{lat} = m_\text{lat}'$, $\chi_n = \chi_n'$ gives a model written in terms of the primed variables in which the fermions have charge $q$.  In the charge-$q$ model, $\theta'$ has periodicity $2 \pi$, which means that, in the original unit charge theory, the values of $\theta$ that differ by $2 \pi / q$ are now considered to be different ``universes'' that are part of the same theory.  As we will explain (see Appendix~\ref{sec:HamiltonianAppendix}), the charge $q$ model possesses a $\Z_q$ lattice one-form symmetry.  When $q$ is even, the lattice translation by one site is a true symmetry of the charge-$q$ massless theory because it changes $\theta'$ by a multiple of $2 \pi$.

While the papers using the MPS, DMRG, and the experimental realizations of the Schwinger model usually consider lattices with open boundary conditions (OBC), we will mostly work with periodic boundary conditions (PBC). The latter have the advantage of realizing the discrete translation symmetries which play an important role in our work, but they also require inclusion of the global $U(1)$ degree of freedom. We will diagonalize our finite size lattice Hamiltonian, and then extrapolate the results to the continuum limit. We will show that accounting for the mass shift (\ref{mLat}) in the lattice Hamiltonian greatly improves the numerical results both for PBC and OBC, and we believe that the MPS, DMRG, and experimental approaches would benefit from including this shift as well.

The rest of this paper is organized as follows.  In Section~\ref{sec:review}, we begin with a brief review of the Schwinger model in the continuum limit, highlighting various results that will later be compared with our numerical studies.  Additional details on the continuum Schwinger model are presented in Appendix~\ref{sec:theta}.  In Section~\ref{sec:lattice}, we discuss the lattice gauge theory \eqref{eq:HLattice}, explain our mass identification \eqref{mLat}, and provide a few results in the strong coupling expansion.  Additional details on the lattice model can be found in Appendix~\ref{sec:HamiltonianAppendix}.  In Section~\ref{sec:numerics}, we present numerical results for various properties of the lattice model \eqref{eq:HLattice} and compare to the continuum limit.  We end with a discussion of our results in Section~\ref{sec:discussion}.

\section{Review of the Schwinger model on a circle}\label{sec:review}

Before studying the lattice Hamiltonian \eqref{eq:HLattice}, let us mention briefly a few results that can be obtained in the continuum limit $a \to 0$ with $Na = L$ kept fixed, without making reference to the lattice description.  The continuum theory is the Schwinger model, with Lagrangian density (for conventions and Hamiltonian formulation, see Appendix~\ref{sec:theta})
\es{Schwinger}{
	{\cal L} = - \frac 1{4e^2} F_{\mu\nu} F^{\mu\nu} - \frac{\theta}{2\pi} \epsilon^{\mu\nu} F_{\mu\nu} 
	+ \bar \Psi ( i \slashed{\partial} - \slashed{A} - m ) \Psi 
}
defined on a circle of circumference $L$.  For a proper definition, one should impose boundary conditions.  While the gauge field should obey periodic boundary conditions (up to gauge transformations), for the fermions one may impose a boundary condition of the form $\Psi(x + L) = e^{i \phi} \Psi(x)$ for some phase $\phi$.  It turns out that all  physical quantities are independent of the phase $\phi$ \cite{Hetrick:1988yg}.  The Schwinger model \eqref{Schwinger} on a circle is analytically solvable when $m=0$ \cite{Lowenstein:1971fc,Casher:1974vf, Coleman:1975pw}, when $m/e \gg 1$ \cite{Coleman:1976uz}, or when $e L \ll 1$ \cite{Shifman91}.

When $m=0$,  as can be seen from bosonization, \eqref{Schwinger} describes a free Schwinger boson of mass $M_S \equiv e / \sqrt{\pi}$, obeying periodic boundary conditions. The single particle dispersion relation is given by the relativistic energy formula
\es{SinglePart}{
	E_k = \sqrt{M_S^2 + p_k^2} \,, \qquad p_k = \frac{2 \pi k}{L} \,,
} 
with quantized momentum $p_k$ and energy $E_k$.  Due to the Schwinger anomaly, the theory is independent of $\theta$, as we review in Appendix~\ref{sec:theta}\@.  As $e L \to 0$, all non-zero momentum states have very large energies, and the remaining spectrum consists of equally spaced energy levels $n M_S$, corresponding to $n$ Schwinger bosons at rest.  The same spectrum can be interpreted as arising from a harmonic oscillator effective potential for the holonomy variable, where this potential is obtained by integrating out the fermions.

For $m \neq 0$, the gap $M_S(m, L)$ (lowest particle mass) and the effective number of particles depend non-trivially on $m$ and $\theta$ \cite{Coleman:1976uz}.  For fixed $m$ and $\theta$, $M_S$ varies from $M_S(m, 0) = M_S$ at $L=0$ to an asymptotic value $M_S(m, \infty)$ attained in the $L \to \infty$ limit.  In \cite{Adam_1997}, this asymptotic value is given in mass perturbation theory as
\begin{equation}\label{eq:mass_gap_perturbative}
	\begin{split}
		M_S^2(m, \infty) &= M_S^2 + 2e^{\gamma} m M_S \cos\theta \\
		&\quad+ e^{2\gamma} m^2\left(A + B\cos 2\theta\right) + \mathcal{O}\left(m^3\right) \,,
	\end{split}
\end{equation}
where $A \approx 1.7277$ and $B\approx -0.6599$.

Interestingly, in the infinite volume limit, the $(m, \theta)$ phase diagram contains a line of first order phase transitions at $\theta = \pi$ and $m> m_c$, where $m_c$ is the critical coupling corresponding to a second order point \cite{Coleman:1976uz}.  The theory goes from one vacuum with no symmetry breaking at small $m/e$ to two vacua with spontaneous symmetry breaking at large $m/e$.  For $m/e \gg 1$, one can understand the two vacua as the configurations where the effective electric field (defined as in \eqref{Eeff}) is $+1/2$ and $-1/2$, and the electrons of mass $m$ are the domain walls between the two vacua.  The critical coupling $m_c$ can be determined, for instance, from the requirement that $M_S(m_c, \infty) = 0$.  The critical exponents $\nu$ and $\beta$ were computed using various methods, the most precise being the DMRG studies of \cite{Byrnes:2002gj,Byrnes:2002nv}, which gave $\nu = 0.99(1)$ and $\beta/\nu = 0.125(5)$.   These values suggest that the critical point is in the 2d Ising universality class with $\nu = 1$ and $\beta = 1/8$.  Refs.~\cite{Byrnes:2002gj,Byrnes:2002nv} found $m_\text{cr} / e = 0.3335(2)$.

Another quantity of interest is the chiral condensate $\langle \bar \Psi \Psi \rangle$.  At $m=0$, it was computed in \cite{Hetrick:1988yg} as a function of $L$:
\es{condensate}{
	\langle \bar \Psi \Psi \rangle_L = - \frac{e^\gamma \cos \theta}{2 \pi^{3/2}}  
	\exp \left( 2 \int_0^\infty \frac{dx}{1 - e^{\mu L \cosh x}} \right) \,.
}
The same result can be interpreted as the chiral condensate at temperature $T = 1/L$ for the theory on an infinite line \cite{Banuls:2016lkq}.  For small circles we have $\langle \bar \Psi \Psi \rangle_0 = 0$, while for large circles, the condensate approaches 
\es{condensateInf}{
	\langle \bar \Psi \Psi \rangle_\infty  = - \frac{e^\gamma}{2 \pi^{3/2}} e  \cos \theta  \approx -0.160\,  e \cos \theta \,.
}
When $m \neq 0$, the condensate $\langle \bar \Psi \Psi \rangle_L$ can no longer be computed analytically, but the expectation is that it is approximated by the formula \eqref{condensate} at small $L$, while at large $L$ it approaches an asymptotic value that depends on $m$ and $\theta$.

\section{Lattice formulation}\label{sec:lattice}

\subsection{Gauss's law and states}

Let us now turn to the study of the lattice Hamiltonian \eqref{eq:HLattice}.
The (anti-)commutation relations obeyed by the operators appearing in \eqref{eq:HLattice} are
\es{CommutLattice}{
	[L_n, U_m] &= \delta_{nm} U_n \,, \qquad [L_n, U_m^\dagger] = -\delta_{nm} U_n^\dagger \,, \\
	\{ \chi_n, \chi_m^\dag \} &= \delta_{nm} \,,
}
with all other (anti-)commutators not explicitly written vanishing.  It is understood that the Hamiltonian acts on a Hilbert space with a vacuum state $\ket{\text{vac}}$ that is annihilated by all $L_n$ and $\chi_n$.

In explicit computations, it is useful to consider a basis of simultaneous eigenstates of the $L_n$ operators and the occupation number operators $N_n = \chi_n^\dagger \chi_n$.  One starts with the Fock vacuum $\ket{\text{vac}}$, which by assumption has $L_n = N_n= 0$ for all $n$.   On it, we can act with $\chi_n^\dagger$, $U_n$ (or $U_n^\dagger = U_n^{-1}$) to construct basis states
\es{Basis}{
	&\ket{n_0, n_1, \ldots n_{N-1}} \ket{\ell_0, \ell_1, \ldots, \ell_{N-1}} \\
	&\qquad \qquad\qquad \qquad  = \prod_{m=0}^{N-1} (\chi_m^\dagger)^{n_m} U_n^{\ell_m} \ket{\text{vac}} 
}
with $L_m = \ell_m$ and $N_m = n_m \in \{0, 1\}$.  As mentioned in the Introduction, gauge invariance requires the Gauss law \eqref{Gauss} and that $L_n \in \Z$ (see Appendix~\ref{sec:HamiltonianAppendix} for a detailed explanation), which means that out of all the states of the form \eqref{Basis}, only those with 
\es{GaugeCond}{
	\ell_m - \ell_{m-1} = n_m - \frac{1 - (-1)^m}{2}  \quad \text{and} \quad \ell_m \in \Z
}
are physical.  In particular, note that with periodic boundary conditions $\ell_{N} = \ell_0$, so adding up all the constraints in \eqref{GaugeCond} one obtains the half-filling condition $\sum_{m=0}^{N-1} n_m = \frac{N}{2}$.

\subsection{Enhanced $\Z_2$ chiral symmetry}\label{sec:mass_shift}

Let us now justify the proposed identification \eqref{mLat} between the lattice mass $m_\text{lat}$ and the continuum parameter $m$.  As explained in detail in Appendix~\ref{sec:theta}, in the continuum Schwinger model at $m=0$, the Schwinger anomaly implies that the Hamiltonians $H_\theta$ and $H_{\theta'}$ are unitarily equivalent for any $\theta$ and $\theta'$.  In particular, one can show that there exists a family of unitary operators ${\cal V}_\alpha$ such that ${\cal V}_\alpha H_\theta {\cal V}_\alpha^{-1} = H_{\theta - 2 \alpha}$.  We now show that at the special value $m_\text{lat} = - \frac{e^2 a}{8}$ the transformation that translates the lattice by one site maps $H_\theta$ to $H_{\theta + \pi}$ in an analogous way.  

Indeed, let us consider the unitary operator ${\cal V}$ that implements this translation by one site, namely
\es{Translation}{
	{\cal V} \chi_n {\cal V}^{-1} &= \chi_{n+1} \,, \qquad {\cal V} U_n {\cal V}^{-1} = U_{n+1}  \,.
}
It follows that the charges $Q_n$ on each link defined in \eqref{Gauss} transform to
\es{LinkCharge}{
	{\cal V} Q_n {\cal V}^{-1} = Q_{n+1} + (-1)^n \,.
}
In order for the Gauss law \eqref{Gauss} to be obeyed, the flux operators $L_n$ should transform as
\es{LnChange}{
	{\cal V} L_n {\cal V}^{-1} = L_{n+1} + \frac{1 + (-1)^n}{2} + \ell \,
}
where $\ell$ is a constant.  We need to take $\ell \in \Z$ so that the eigenvalues of the $L_n$ operators are still integers after the one-site translation, and we can choose $\ell =0$ without loss of generality.   An explicit computation using \eqref{Translation}, \eqref{LnChange}, and the Hamiltonian \eqref{eq:HLattice} shows that
\es{ChangeH}{
	{\cal V} H_\theta {\cal V}^{-1}
	&= H_{\theta +  \pi} - 2 m_\text{lat} \sum_{n=0}^{N-1} (-1)^{n} \chi^\dag_n \chi_n \\
	&{}+ \frac{e^2 a}{2} \sum_{n=0}^{N-1} \left( (-1)^n L_{n+1} + \frac 14 \right) \,.
}
Using the Gauss law \eqref{Gauss}, we see that the last two terms precisely cancel provided that 
$m_\text{lat}$ is taken to be 
\es{specialPoint}{
	m_\text{lat}  = -\frac{e^2 a}{8} \,.
}
Thus, we have shown that at this special value of $m_\text{lat}$,  ${\cal V} H_\theta {\cal V}^{-1} = H_{\theta + \pi}$ on the gauge-invariant states,  so the Hamiltonians $H_\theta$ and $H_{\theta + \pi}$ are unitarily equivalent.  This should be interpreted as the discretized analog of the fact that in the continuum limit, the Hamiltonians with any two values of $\theta$ are unitarily equivalent, as shown in Appendix~\ref{sec:theta}\@.  In particular, the operator ${\cal V}$ here is the discretized analog of ${\cal V}_{-\pi/2}$ defined in \eqref{ValphaDef}.

Since the value \eqref{specialPoint} should be interpreted as the massless point of the lattice model, the identification in \eqref{mLat} between the continuum mass $m$ and the lattice mass $m_\text{lat}$ follows. This choice yields the same continuum limit at $m = m_\text{lat}$, but, as we show in Section \ref{sec:numerics}, the choice \eqref{mLat} leads to a far faster convergence towards the continuum as a function of $N$.

\subsection{Strong coupling expansions}\label{sec:strong_coupling}

One of the first uses of the
lattice Hamiltonian approach to the Schwinger model \cite{Banks:1975gq} was for the strong coupling expansions in powers of $y= 1/(e a)^4$, which can be developed directly in the $N\rightarrow \infty$ limit. 
In \cite{Banks:1975gq}, the expansions were performed to order $y^2$ with coefficients that are functions of $\mu= 2 m_{\rm lat}/(e^2 a)$, but the most extensive such results to date may be found in \cite{Hamer:1997dx}. For example, the mass gap $ E_1- E_0 = \frac{e^2 a}{2} (\omega_1- \omega_0)$, is derived from
\es{gapexp}{
	\omega_1- \omega_0 = \delta\omega &=  1+ 2 \mu + \frac{2y}{1+ 2 \mu} - \frac{2 (5+ 2 \mu)y^2 }{(1+ 2 \mu)^3} \notag \\
	& +  \frac{4 (59+ 68 \mu + 24\mu^2 + 4\mu^3 )y^3 }{(1+ 2 \mu)^5 (3+ 2\mu)} + O(y^4)
	\ .}
In order to describe the theory with discrete chiral symmetry, we set $\mu=-1/4$ to obtain
\be
\label{gapextra}
\delta \omega=  \frac{1}{2} + 4y - 72 y^2  +2224 y^3 + O(y^4)\ .
\ee  
To order $y^2$ this agrees with \cite{Berruto:1997jv}.
We extrapolate (\ref{gapextra}) to large $y$ assuming the asymptotic behavior $\sim y^{1/4}$ required by the existence of the continuum limit. This is accomplished by applying the $(2,1)$ Pad\' e approximant to $(\delta \omega)^4$, which produces a smooth function $\delta\omega (y)$:
\es{Pade}{
	\delta \omega(y) \approx \left( \frac{\frac{1}{16} +\frac{95 y}{24} + \frac{152 y^2}{3} }{1 + \frac{94 y}{3}} \right)^{\frac 14}\,.
} 
This method of extrapolation leads to the estimate 
\be E_1- E_0\approx \left (\frac{19}{188}\right )^{1/4}e\approx 0.56383 e\ ,
\ee
which is very close to the exact continuum result $M_S= e/\sqrt{\pi} \approx 0.56419 e$.
The $0.06 \%$ agreement of our extrapolation with the continuum limit is much better than what was obtained in \cite{Hamer:1997dx} for $\mu=0$. This illustrates the importance of using the Hamiltonian with discrete chiral symmetry to describe the massless Schwinger model. Closely related results may be found in \cite{Berruto:1997jv}, although the agreement with the continuum limit found there was not as good.

To obtain the strong coupling expansion of the ground state expectation value of $\bar \Psi \Psi$, we may differentiate with respect to $m$ the expression for the ground state energy:
\be
\langle  \bar \Psi \Psi \rangle = \frac{1}{Na} \frac{\partial E_0}{\partial m}= \frac{1}{N a} \frac{\partial \omega_0}{\partial \mu}
\ee
Evaluating this at the chirally symmetric point $\mu=-1/4$, we obtain from \cite{Hamer:1997dx}

	\es{VEVExpanded}{
		-a \langle  \bar \Psi \Psi \rangle = \frac{1}{2}- 8 y + 288 y^2 - \frac{306688}{25} y^3 + O(y^4)
	}
	The coefficient of $y^2$ does not agree with \cite{Berruto:1997jv} (we have checked some of the strong coupling results in \cite{Hamer:1997dx}).
	Using the results up to $O(y^5)$ \cite{Hamer:1997dx} we extrapolate to large $y$ assuming the asymptotic behavior $\sim y^{-1/4}$ required by the existence of the continuum limit.  The $(2, 3)$ Pad\'e approximant of \eqref{VEVExpanded} raised to the fourth power gives the continuum estimate
	$\langle  \bar \Psi \Psi \rangle \approx -0.164 e$, which is close to the exact result \eqref{condensateInf} in the massless Schwinger model. 

\section{Numerical results}\label{sec:numerics}

We now discuss the results we obtain by numerical diagonalization of \eqref{eq:HLattice}. In particular, we exhibit the improvement in the convergence of the results to the continuum limit as a result of the mass shift discussed in Section \ref{sec:mass_shift}.

In order to make \eqref{eq:HLattice} into a finite problem, we first have to truncate the infinite basis of states \eqref{eq:states}. For a lattice of $N$ sites, there are $\binom{N}{N/2}$ ways to assign the occupation numbers $n_m$. For each of these, the average electric field 
\es{AvE}{
	{\cal E} \equiv \frac{1}{N} \sum_{n=0}^{N-1} L_n 
}
is $\mathcal{E} \in \mathcal{E}_0 + \mathbb{Z}$, where $\mathcal{E}_0$ is determined by the occupation numbers as
\begin{equation}
	\mathcal{E}_0 \equiv - \frac{1}{N}\sum_{n=0}^{N-1} n Q_n \pmod{1} \,.
\end{equation}
(See also the discussion that leads to \eqref{InvLargeCalE} in Appendix~\ref{sec:HamiltonianAppendix}.) Since the Hamiltonian contains a term proportional to $\mathcal{E}^2$ (see \eqref{HLattice3}), we can get a very good approximation to the low-lying spectrum by keeping only the few lowest-magnitude values of $\mathcal{E}$. In fact, as we increase the truncation of possible $\mathcal{E}$ values the low-lying spectrum converges exponentially fast, and we are able to take a fixed truncation without any meaningful loss in accuracy of the results \footnote{For instance, on a lattice of $N = 16$ sites with a circle length $Le = 1$, increasing the truncation from $\mathcal{E}\in[-4.5,4.5]$ to $\mathcal{E}\in[-5.5,5.5]$ does not change the mass gap up to machine precision, and the relative change in $\langle \bar\Psi \Psi\rangle$ is less than $10^{-5}$.}. For all the plots in this paper, we truncate the possible values to $\mathcal{E}\in [-5.5, 5.5]$. We consider lattices of up to $N = 16$ sites, which corresponds to a basis of $11\binom{16}{8} \approx 1.4\times 10^5$ states. We use SLEPc to diagonalize these large matrices \cite{petsc-user-ref,petsc-efficient,slepc-toms,slepc-manual}, and a single diagonalization for 16 sites typically takes a few CPU minutes. The majority of our results are already well-converged at $N = 10$, with a basis of 2772 states, for which it is perfectly feasible to perform the diagonalizations on a laptop computer.

To compare with known exact and perturbative results, we will study three observables: the mass gap, the chiral condensate $\langle \bar\Psi \Psi\rangle$, and the electric field density $\langle \mathcal{E}\rangle$.

\subsection{Mass gap}

\begin{figure}
	\centering
	\includegraphics[width=\linewidth]{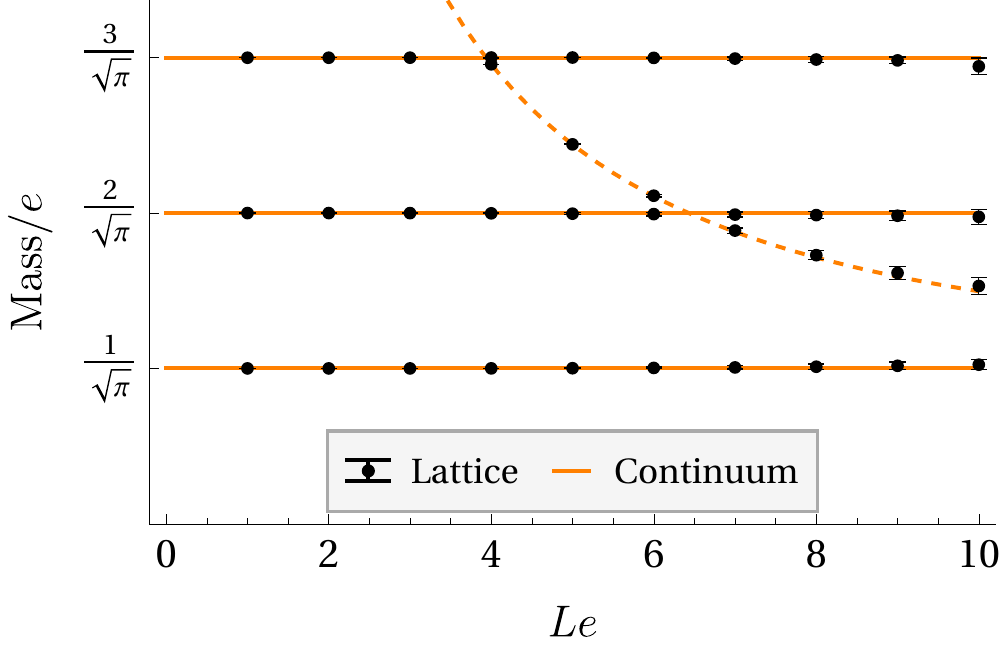}
	\caption{The spectrum of excitations above the ground state for the massless Schwinger model as a function of $L$. We reproduce the Schwinger mass $M_S = \frac{e}{\sqrt{\pi}}$ and observe the beginning of a tower of multi-particle states, along with a single-particle state with nonzero momentum obeying \eqref{SinglePart} (orange dashed). For each point, we compute the mass at $N = 10,12,14,16$ and then extrapolate to $N\to\infty$ and plot $1\sigma$ confidence intervals.}
	\label{fig:spectrum}
\end{figure}

Figure \ref{fig:spectrum} shows the spectrum of excitations above the ground state in the model with $m = 0$ as a function of $L$. For each point, we compute the mass at $N = 10$ through $N = 16$ and then extrapolate to $N\to\infty$ and plot $1\sigma$ confidence intervals. We can clearly see a tower of $n$-particle states of zero momentum %
and a one-particle state of minimal nonzero momentum. 

\begin{figure}
	\centering
	\includegraphics[width=\linewidth]{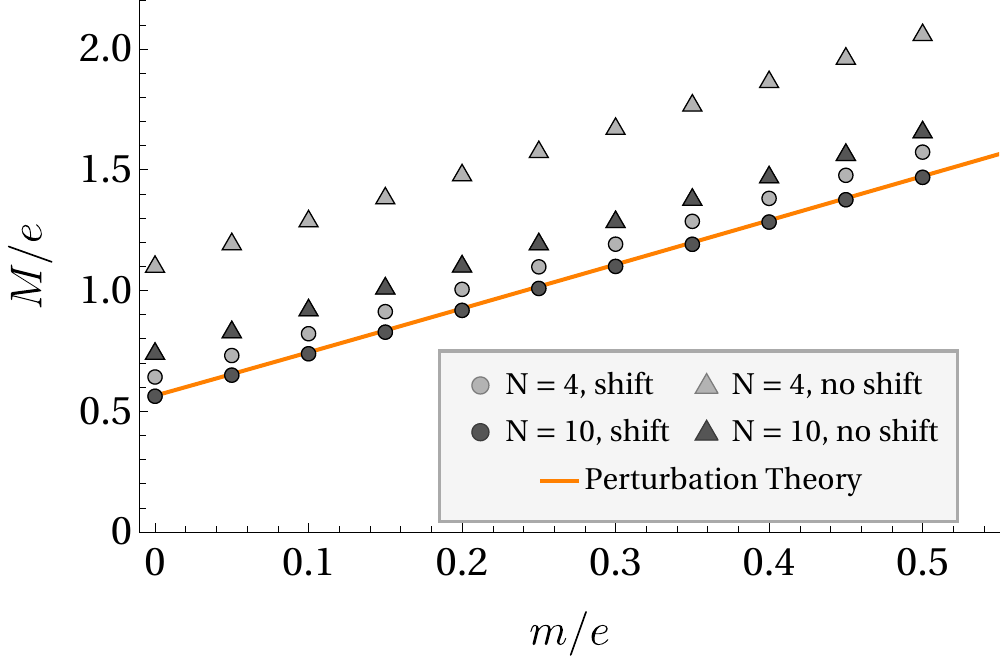}
	\caption{The mass gap as a function of $m/e$ at $\theta = 0$ and $Le = 8$. We show numerical results using $N = 4$ and $N = 10$ lattice sites, both with and without the mass shift. We compare with the perturbative expression \eqref{eq:mass_gap_perturbative} given in \cite{Adam_1997}, showing that the numerics with the mass shift converge much more quickly.}
	\label{fig:mass_gap_finiteN}
\end{figure}

In Figure \ref{fig:mass_gap_finiteN}, we show the close agreement between our numerical results for the mass gap $M_S$ and the $L = \infty$ value \eqref{eq:mass_gap_perturbative} at $\theta = 0$ and small $m$. We work at fixed $Le = 8$, for which the mass gap has nearly converged to its $L\to\infty$ limit. Even with only ten sites, we find remarkably good agreement between our numerical results and perturbation theory. We also show the results we would obtain without using the mass shift introduced in this paper, which converge much more slowly to the continuum limit.

\subsection{Chiral condensate}

\begin{figure}
	\centering
	\includegraphics[width=\linewidth]{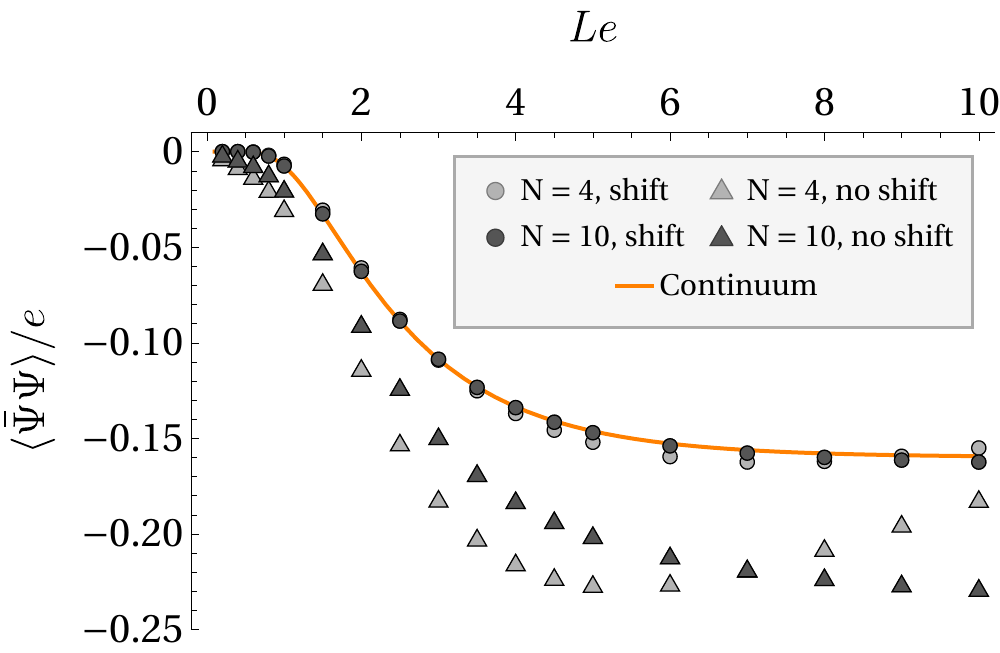}
	\caption{The chiral condensate as a function of $L$ with $m = \theta = 0$. We show numerical results using $N = 4$ and $N = 10$ lattice sites, and both with and without the mass shift. We compare with the exact result \eqref{condensate}, showing that the numerics with the mass shift converge much more quickly.}
	\label{fig:condensate}
\end{figure}

We can compute the chiral condensate $\langle \bar\Psi\Psi\rangle$ numerically as the expectation value of the mass operator in the ground state. In Figure \ref{fig:condensate}, we plot the exact value of the condensate as a function of $L$ along with our numerical results. For the numerics, we give examples with $N = 4$ and $N = 10$ lattice sites, and both with and without the mass shift. Again, we find that the mass shift dramatically improves convergence towards the exact result.

\begin{figure}
	\centering
	\includegraphics[width=\linewidth]{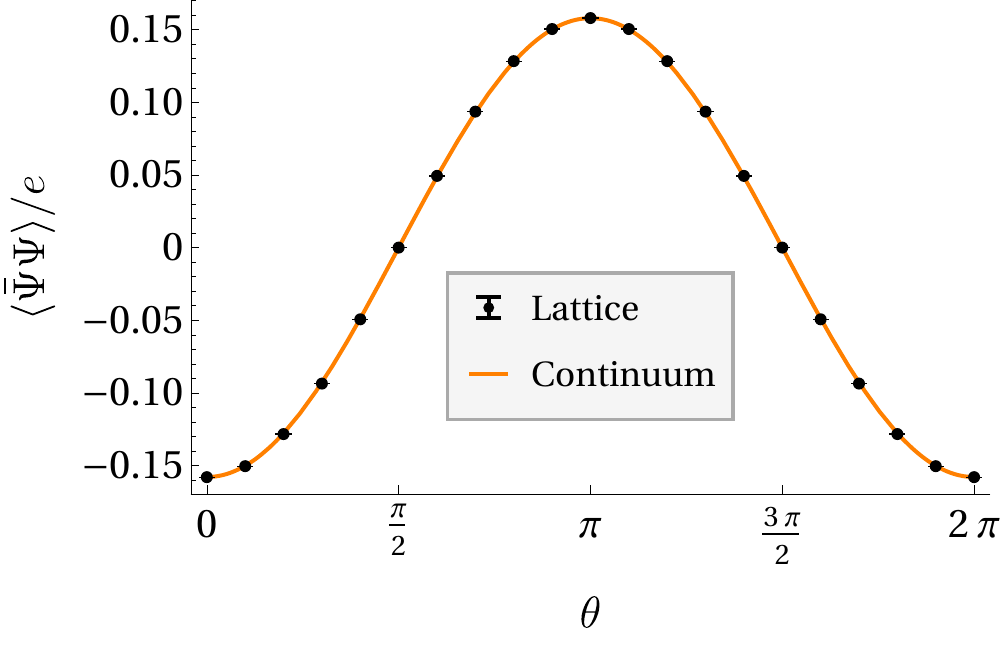}
	\caption{The chiral condensate as a function of $\theta$ with $m = 0$ and $Le = 8$. The 1$\sigma$ confidence intervals plotted come from extrapolating the values at $N=10,12,14,16$ to the continuum limit $N\to\infty$. The numerical results exhibit the $\cos\theta$ dependence in \eqref{condensate}.}
	\label{fig:condensate_theta}
\end{figure}

In Figure \ref{fig:condensate_theta}, we fix $Le = 8$ and plot $\langle\bar\Psi\Psi\rangle$ as a function of $\theta$. We see that our numerics, extrapolated to large $N$ as in Figure \ref{fig:spectrum}, reproduce the exact $\cos\theta$ dependence in \eqref{condensate} extremely well. In particular, even on a finite lattice we have $\langle \bar\Psi\Psi\rangle(\theta + \pi) = -\langle\bar\Psi\Psi\rangle(\theta)$, as a consequence of the discrete chiral symmetry that maps $\theta\mapsto\theta+\pi$. This property would not hold on a finite lattice without the shifted definition of $m_\text{lat}$.

\subsection{Average electric field}

Another quantity of interest is the average electric field condensate, $\langle\mathcal{E}\rangle$.  At $m = 0$ and $\theta = 0$, the ground state has zero electric field on every link, and so $\langle\mathcal{E}\rangle = 0$. Since the massless Schwinger model is independent of $\theta$ (see Appendix \ref{sec:theta}), we have $\langle\mathcal{E}\rangle = 0$ for $m = 0$ and any $\theta$. The value of $\langle\mathcal{E}\rangle$ is given in mass perturbation theory in \cite{Adam_1997} as
\begin{equation}
	\langle \mathcal{E}\rangle/e = \frac{e^\gamma}{\sqrt{\pi}} \frac{m}{e} \sin(\theta) + \frac{e^{2\gamma}}{4\pi} E_+ \left(\frac{m}{e}\right)^2 \sin(2\theta) + \mathcal{O}\left(m^3\right),
\end{equation}
where $E_+ \approx -8.9139$.

\begin{figure}
	\centering
	\includegraphics[width=\linewidth]{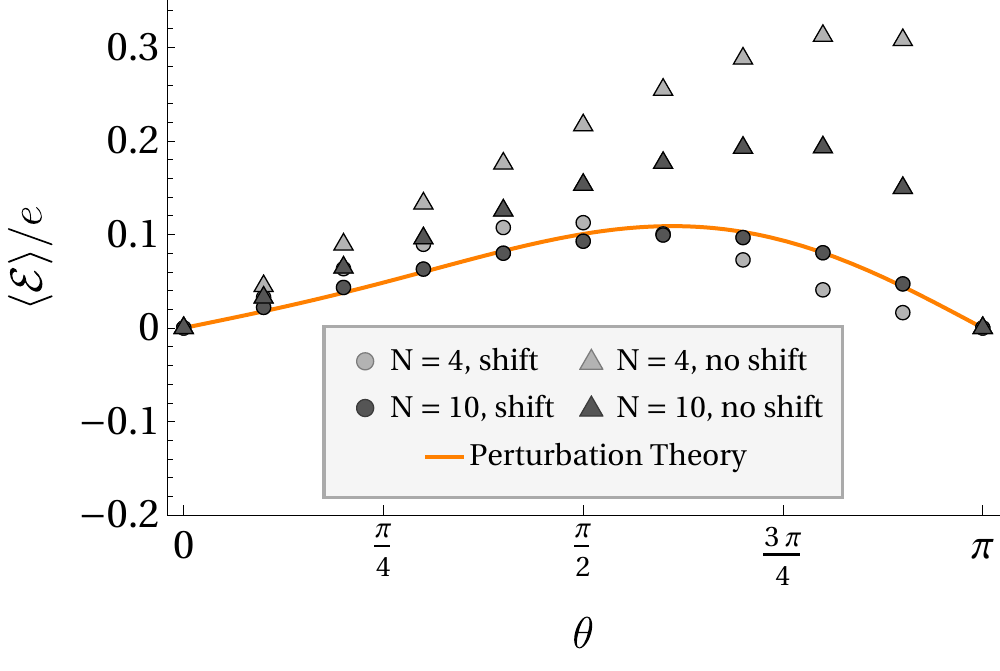}
	\caption{The electric field condensate as a function of $\theta$ for $m/e = 0.1$ and $Le = 8$. We show numerical results using $N = 4$ and $N = 10$ lattice sites, and both with and without the mass shift. We compare with the perturbative expansion given in \cite{Adam_1997}, showing that the numerics with the mass shift converge much more quickly.}
	\label{fig:avgE}
\end{figure}

In Figure \ref{fig:avgE}, we plot the electric field condensate as a function of $\theta$ at $m/e = 0.1$ and $Le = 8$ (large enough that the results are nearly converged to their $L\to\infty$ limit). Again we see that the numerics with the mass shift converge much more quickly.

\subsection{Phase Transition}

The critical mass $m_c$ for the phase transition discussed in the introduction has been computed before via finite-size scaling on the lattice, without the mass shift introduced in this paper \cite{Hamer:1982mx}. Their approach is to first fix the lattice spacing $a$, and calculate the scaled mass gap ratio as a function of the mass,
\begin{equation}
	\alpha(N,m;a) \equiv \frac{N M(N, m;a)}{(N+2)M(N+2,m;a)} \,.
\end{equation}
They then determine the masses $m_*(N,a)$ for which $\alpha(N, m_*(N, a); a) = 1$, and extrapolate to $m_*(\infty, a)$. Finally, by taking $a\to 0$, they recover an estimate of the critical mass in the continuum limit. This method is discussed in more detail in \cite{Hamer_1980}.

\begin{figure}
	\centering
	\includegraphics[width=\linewidth]{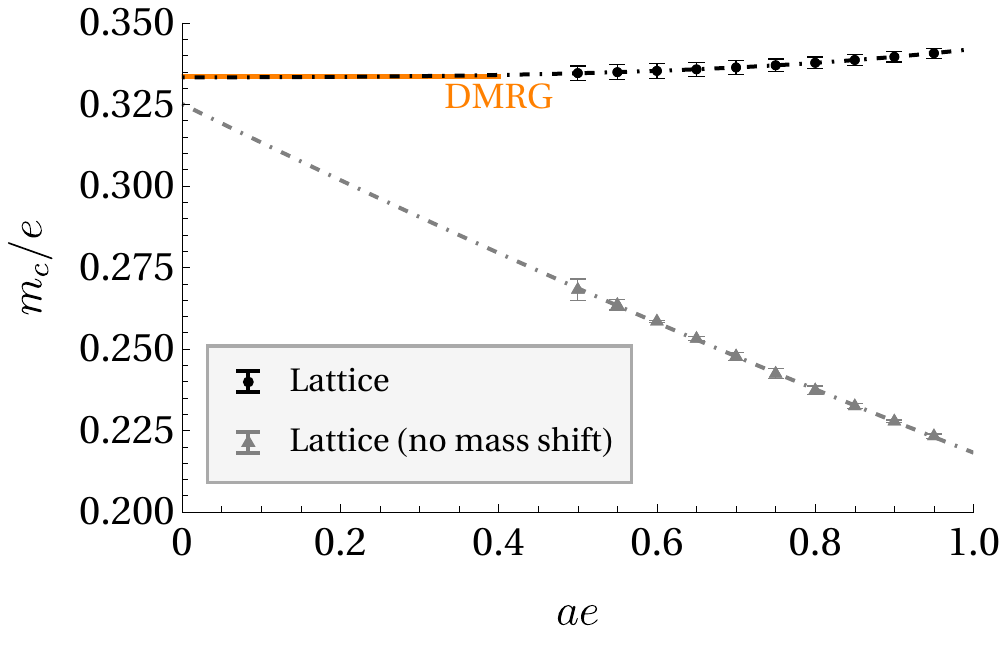}
	\includegraphics[width=\linewidth]{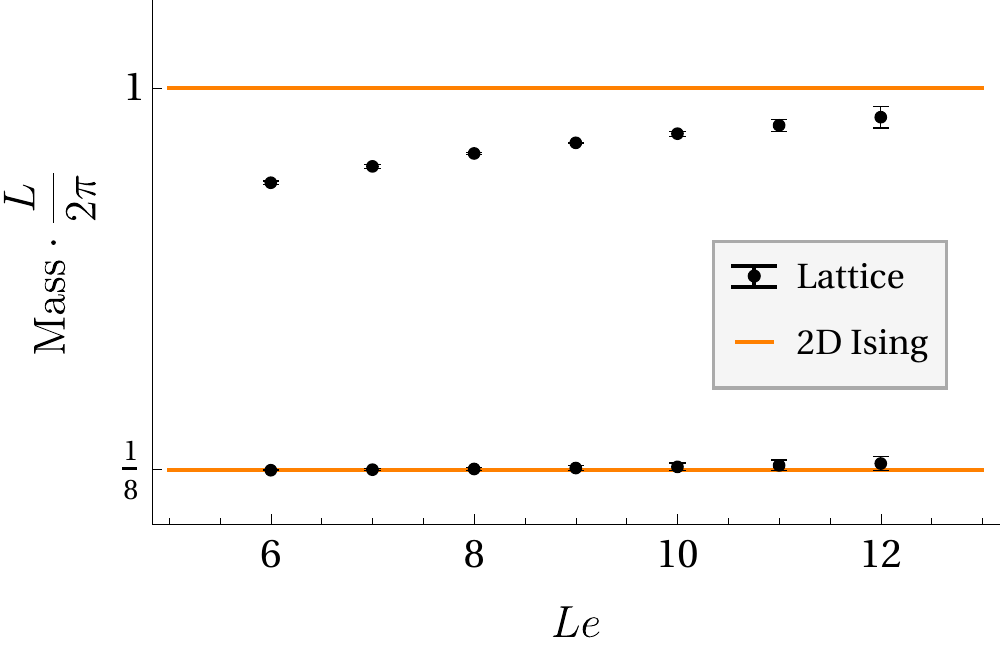}
	\caption{(Top) The finite-size scaling estimation of the critical mass $m_c$, shown both with and without the mass shift introduced in this paper. Without the mass shift, we recover from a linear extrapolation the result of \cite{Hamer:1982mx}, $m_c/e = 0.325 \pm 0.02$. With the mass shift, the linear term in $a$ appears to go to zero, and by fitting a model of the form $c_0 + c_2 (ea)^2 + c_4 (ea)^4$ we find $m_c/e = 0.333\pm 0.005$. (Bottom) We compute the lowest two translationally invariant states in the model at the critical mass in units of the inverse circle radius, and compare with the lowest two operators in the 2D Ising CFT\@. Extrapolating the lattice results to $L\to\infty$ gives $\Delta_\sigma = 0.14(4)$ and $\Delta_\epsilon = 1.09(11)$, consistent with the 2D Ising values $\Delta_\sigma = 1/8$ and $\Delta_\epsilon = 1$.}
	\label{fig:transition}
\end{figure}

We repeat their analysis using our shifted lattice mass. Figure~\ref{fig:transition} shows the significantly improved convergence as a function of the lattice spacing $a$. By extrapolating our results to the $a\to 0$ continuum limit, we find
\begin{equation}
	m_c/e = 0.333 \pm 0.005 \,, 
\end{equation}
consistent with the value $m_c/e = 0.3335 \pm 0.0002$ obtained from DMRG studies \cite{Byrnes:2002gj,Byrnes:2002nv}.

We can also estimate the scaling dimensions of operators in the 2D Ising CFT at the critical point by calculating the translationally-invariant spectrum for $m = m_c$ in units of the inverse radius of the circle.
This statement is true in the $L\to \infty$ limit, where the theory reaches the true IR limit. 
At finite $L$, we see good agreement with the dimension of the lowest operator in the 2D Ising model, $\Delta_\sigma = 1/8$, and consistency with the second-lowest operator at $\Delta_\epsilon = 1$. The critical exponents are derived from these scaling dimensions as
\begin{equation}
	\beta = \frac{\Delta_\sigma}{2-\Delta_\epsilon} = \frac{1}{8} \,, \qquad \nu = \frac{1}{2-\Delta_\epsilon} = 1 \,.
\end{equation}

\subsection{Open Boundary Conditions}

We have shown that the mass shift (\ref{mLat}) greatly improves the rate of convergence to the continuum limit in the lattice Hamiltonian approach \cite{Banks:1975gq,Hamer:1997dx} to the Schwinger model. This is true both for the numerical results with periodic boundary conditions on finite lattices and for the strong coupling expansions on the infinite lattice. One may also wonder to what extent the mass shift improves the results for the open boundary conditions (OBC) with the numerically accessible values of lattice size $N$.  A simplification for the OBC is that there is no longer a bosonic $U(1)$ rotator variable, and the model simply reduces to a fermionic chain with long-range interaction. On the one hand, the OBC certainly violate the lattice translation symmetry that realize the discrete chiral symmetry of the massless theory; on the other, for sufficiently large $N$, the theory should be close to the infinite lattice limit where the discrete chiral symmetry singles out the massless Schwinger model.

\begin{figure}
	\centering
	\includegraphics[width=\linewidth]{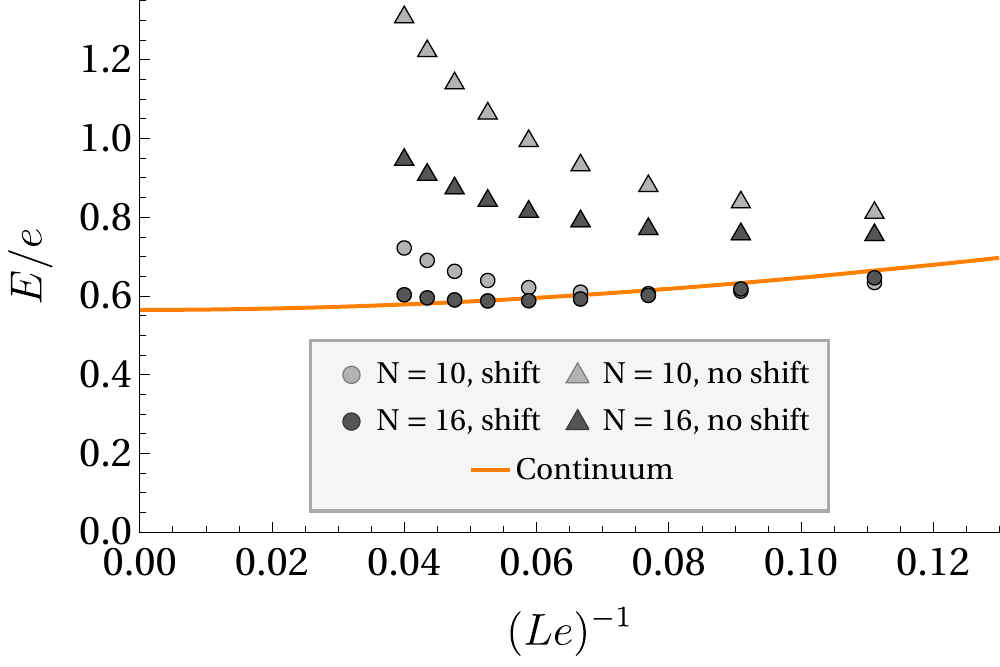}
	\includegraphics[width=\linewidth]{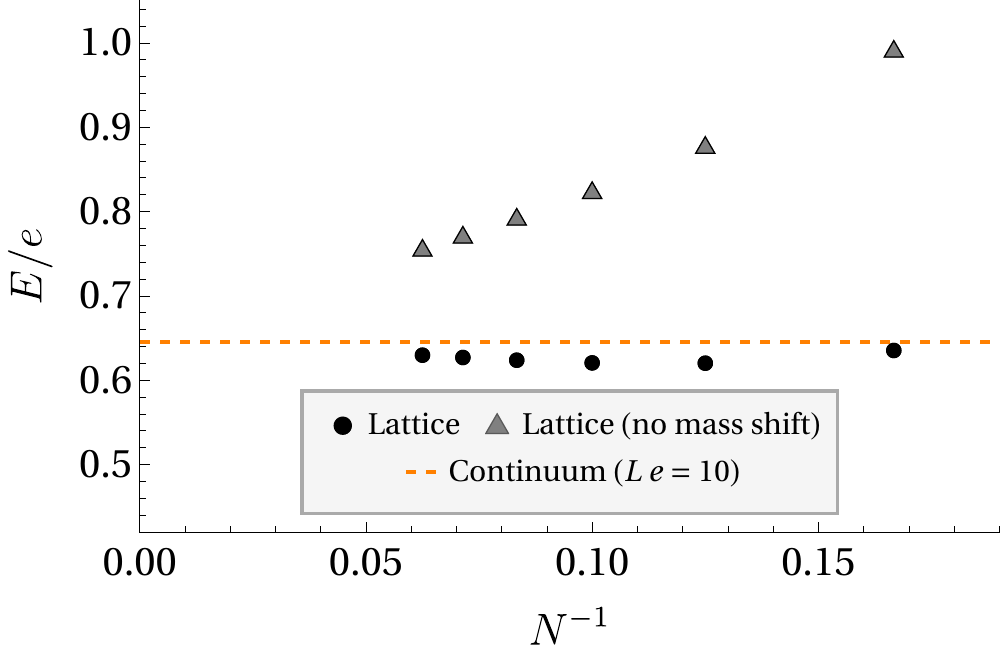}
	\caption{(Top) The lowest energy excitation as a function of $L$ at $\theta = 0$ and $m = 0$, on a lattice with open boundary conditions with $N = 10$ and $N = 16$ lattice sites both with and without the mass shift \eqref{mLat}. Including the mass shift significantly improves the convergence to \eqref{eq:obc_mass_gap} as a function of $N$ also in the case of open boundary conditions. (Bottom) A more detailed look at the improved convergence in $N$ to the continuum result from \eqref{eq:obc_mass_gap} at fixed $Le = 10$.}
	\label{fig:obc}
\end{figure}

In Figure \ref{fig:obc} we present numerical results for the mass gap with OBC. We show estimates for the mass gap obtained both with and without the mass shift, for $N = 10$ and $N = 16$ lattice sites and at various system sizes $L = Na$. For a fixed system size with open boundary conditions, the wavefunction of the Schwinger boson in the continuum model should vanish at $x = 0$ and $x = L$, so the energy of the lowest state is
\begin{equation}\label{eq:obc_mass_gap}
	E(L) = \sqrt{M_S^2 + \left(\frac{\pi}{L}\right)^2}.
\end{equation}

With the mass shift \eqref{mLat} we observe a significant improvement in the convergence of the lattice results towards \eqref{eq:obc_mass_gap} as a function of $N$. This encourages us to believe that the mass shift with be valuable in the further work on models with OBC,
both numerical and experimental.

\section{Discussion}\label{sec:discussion}

Our results show that the Kogut-Susskind Hamiltonian lattice gauge theory can be a precision numerical tool for studying models in $1+1$ dimensions even before the more advanced methods involving matrix product states and DMRG are applied. In view of these encouraging results, it would be interesting to explore other models using the lattice Hamiltonian approach. For example, in the multi-flavor Schwinger model there is a mass shift analogous to (\ref{mLat}): $m_{\rm lat}= m- N_f e^2 a/8$.
When the number of flavors $N_f$ is odd, then the $m=0$ theory is invariant under the translation by one lattice unit accompanied by the shift of $\theta$ by $\pi$. When $N_f$ is even, then $\theta$ remains unchanged so that the translation by one lattice unit is a true symmetry. 
In particular, the two-flavor model \cite{Coleman:1976uz} should be quite accessible numerically (for recent results using Euclidean lattice theory, see \cite{Albergo:2022qfi}. We also plan to use the Hamiltonian lattice theory to study $SU(N)$ gauge theory coupled to an adjoint Majorana fermion. In this case, already in the 1990s there was a multitude of numerical results available from the light-cone 
Hamiltonian approach \cite{Dalley:1992yy,Bhanot:1993xp,Gross:1997mx}. In recent years there has been renewed interest in adjoint QCD$_2$ 
\cite{Cherman:2019hbq,Komargodski:2020mxz,Dempsey:2021xpf}, and we plan to apply the lattice Hamiltonian approach to models of this type on a spatial circle. 

\section*{Acknowledgments}

This work was supported in part by the US National Science Foundation under Grants No.~PHY-1914860, PHY-2111977, PHY-2209997 and PHY-1748958, and by the Simons Foundation Grants No.~488653 and 917464.  
IRK thanks the Kavli Institute of Theoretical Physics for its hospitality during the workshop ``Confinement, Flux Tubes, and Large $N$," where some of this work was carried out.

\appendix

\section{More details on the continuum Schwinger model}\label{sec:theta}

\subsection{Conventions and Hamiltonian formulation}

When writing the Lagrangian density \eqref{Schwinger}, we used the convention where $(x^0, x^1) = (t, x)$, $\{\gamma^\mu, \gamma^\nu\} = 2 \eta^{\mu\nu} =2 \diag\{1, -1\}$, and $\bar \Psi = \Psi^\dagger \gamma^0$.  We also took $\epsilon^{01} = - \epsilon^{10} = 1$ and $F_{\mu\nu} = \partial_\mu A_\nu - \partial_\nu A_\mu$.  It is useful to also define the chirality matrix $\gamma^5 = \gamma^0 \gamma^1$. 

In the gauge $A_0 = 0$, the Hamiltonian that follows from the Lagrangian density \eqref{Schwinger} is
\es{Hamilt}{
	H &= \int_0^L dx \biggl[\frac{e^2}{2}  \left( E(x) + \frac{\theta}{2 \pi} \right)^2 \\
	&{}- i \Psi^\dagger(x) \gamma^5 \left(\partial_1 + i A_1(x) \right) \Psi(x) + m \Psi^\dagger \gamma^0 \Psi \biggr] \,,
}
where $E(x) \equiv \frac{\partial {\cal L}}{\partial \dot{A}_1}$ is the electric field. The canonical (anti-)commutation relations are
\es{Commut}{
	\{\Psi_\alpha(x), \Psi_\beta^\dagger(y)\} &= \delta_{\alpha\beta} \delta(x - y) \,, \\
	[E(x), A_1(y)] &= -i \delta(x-y) \,,
}
with all other (anti-)commutators vanishing.

The Hamiltonian \eqref{Schwinger} is invariant classically under gauge transformations that act as $\Psi(x) \to e^{i \Lambda(x)} \Psi(x)$ and $A_1(x) \to A_1(x) - \partial_x \Lambda(x)$ for any $\Lambda(x)$ such that the $U(1)$ group element $e^{i \Lambda(x)}$ is well defined on the circle.  Quantum mechanically, this transformation is implemented by conjugation by a unitary operator 
\es{ULambda}{
	{\cal U}_\Lambda = \exp\left[ - i \int dx\, \left(  E(x) \partial_x \Lambda(x) + \rho(x) \Lambda(x) \right) \right] \,,
}
where $\rho(x)$ is the charge density operator
\es{rhoDef}{
	\rho(x) \equiv \Psi^\dagger(x) \Psi(x) \,.
}
Indeed, the commutation relations \eqref{Commut} imply that ${\cal U}_\Lambda \Psi(x) {\cal U}_\Lambda^{-1} = e^{i \Lambda(x)} \Psi(x)$ and ${\cal U}_\Lambda A_1(x) {\cal U}_\Lambda^{-1} = A_1(x) - \partial_x \Lambda(x)$.

As in any gauge theory, the physical (gauge-invariant) states are those for which ${\cal U}_\Lambda = 1$ for any $\Lambda$.  When $\Lambda(x)$ is infinitesimal, this condition implies the Gauss law 
\es{GaussContinuum}{
	\partial_x E(x) = \rho(x)  \,,
}
which should be imposed as a constraint on all physical states.  The Gauss law \eqref{GaussContinuum} is not equivalent to ${\cal U}_\Lambda = 1$ for all $\Lambda$;  one should also impose invariance under ``large'' gauge transformations, for which $\Lambda(x)$ is not a well-defined function on the circle, but $e^{i \Lambda(x)}$ is.  An example is $\Lambda_\text{large}(x) = 2 \pi x / L$.  The transformation corresponding to $\Lambda_\text{large}$ is implemented by the unitary operator
\es{calU}{
	{\cal U}_{\Lambda_\text{large}} = e^{- \frac{2\pi i}{L} \int_0^L dx \, \left( E(x) + x \rho(x) \right)}
	= e^{ - 2 \pi i E(0)}  \,,
}
where the second equality follows from \eqref{GaussContinuum} and integration by parts. Thus, invariance under large gauge transformations, ${\cal U}_{\Lambda_\text{large}} = 1$, is equivalent to the requirement $E(0) \in \Z$.  This, in turn, is equivalent to the condition that $E(x) \in \Z$ for any given $x$.

Note that while the electric field $E(x)$ is always an integer, the parameter $\theta/ 2 \pi$ can be interpreted as a fractional background electric field, as can be seen from the fact that the electric field density in the Hamiltonian \eqref{Hamilt} involves the ``effective'' electric field
\es{Eeff}{
	E_\text{eff}(x) = E(x) + \frac{\theta}{2 \pi} \,.
}

\subsection{Chiral symmetry at $m=0$} 

In the rest of this section, let us set $m = 0$.   When $m=0$, the Hamiltonian \eqref{Hamilt} is invariant classically under an axial transformation under which $ \Psi(x) \to e^{i \alpha \gamma^5} \Psi(x)$, with $\alpha$ being a transformation parameter.  Using Noether's theorem, we can construct the axial current 
\es{AxialCurrent}{
	j^{\mu 5}(x) \equiv \bar \Psi(x) \gamma^\mu \gamma^5 \Psi(x)
	= \Psi^\dagger(x) \gamma^0 \gamma^\mu \gamma^5 \Psi(x)
} 
and the associated axial charge 
\es{Q5Def}{
	Q_5 \equiv \int_0^L dx\, j^{05}(x) \,.
}
Quantum mechanically, the charge $Q_5$ is not conserved (it does not commute with the Hamiltonian) due to the Schwinger anomaly.  The most basic statement of this anomaly is that, while the axial charge density $j^{0 5}(x) = \Psi^\dagger(x) \gamma^5 \Psi(x)$ commutes with $A_1(y)$ and has the appropriate commutation relations with the fermions, 
\es{j5CommFerm}{
	[j^{0 5}(x), A_1(y)] &=0\,, \\
	[j^{0 5}(x), \Psi(y)] &= -\delta(x-y) \gamma^5 \Psi(x)\,, \\
	[j^{0 5}(x) , \Psi^\dagger(y)] &=   \delta(x-y) \Psi^\dagger(x)  \gamma^5 \,,
}
it does not actually commute with the electric field $E(y)$.  This can be seen as a regularization effect:  if we point split the product of operators in \eqref{AxialCurrent} while preserving gauge invariance
\es{AxialChargeSplit}{
	j^{05}_\text{reg}(x) = \Psi^\dagger(x + \epsilon) \gamma^5 \Psi(x) e^{-i \int_x^{x + \epsilon} dx'\, A_1(x') } \,,
}
then, because $A_1(x')$ does not commute with $E(y)$, we would find that 
\es{AxialRegCommut}{
	[j^{05}_\text{reg}(x), E(y)] =  j^{05}_\text{reg}(x) \times 
	\begin{cases} 1 & \text{if $y \in (x, x + \epsilon)$} \,, \\
		0 & \text{otherwise} \,.
	\end{cases}
}
But  $j^{05}_\text{reg}(x)$ itself is a divergent quantity in the $\epsilon \to 0$ limit, where the divergence comes from adding up the axial charge densities of the fermions in the Dirac sea: $ j^{05}_\text{reg}(x) = \frac{i}{\pi \epsilon} + {\cal O}(\epsilon^0)$ \cite{Shifman91}.  Combining this result with \eqref{AxialRegCommut} and taking $\epsilon \to 0$, one obtains the commutator
\es{Commutj5E}{
	[j^{05}(x), E(y)] = \frac{i}{\pi} \delta(x-y) \,.
}
This equation can also be derived from the bosonized description of the model \cite{Manton:1985jm}.  

From \eqref{Commutj5E}, various other facts related to the anomaly can be derived.  For instance, taking the commutator with the Hamiltonian, one obtains
\es{timeDer}{
	i [H, j^{05}(x)] = - \partial_x j^{15} + \frac{e^2}{\pi} \left( E(x) + \frac{\theta}{2 \pi} \right) \,,
}
where $j^{15}(x) = \Psi^\dagger(x) \Psi(x)$.  This equation is nothing but the equation $\partial_\mu j^{\mu 5} =  \frac{e^2}{\pi} \left( E(x) + \frac{\theta}{2 \pi} \right)$ describing the non-conservation of the axial current.

To prove independence of the spectrum on $\theta$, let us define the unitary operator implementing a finite axial transformation with parameter $\alpha$:
\es{ValphaDef}{
	{\cal V}_\alpha \equiv e^{i \alpha Q_5} \,.
} 
From the definition \eqref{Q5Def} and the commutation relations \eqref{j5CommFerm} and \eqref{Commutj5E}, we can then find
\es{Q5H}{
	[Q_5, H] &= \frac{i e^2}{\pi} \int dx\, \left( E(x) + \frac{\theta}{2 \pi} \right) \,, \\
	[Q_5, [Q_5, H]] &= -\frac{e^2}{\pi^2} L \,.
}
Writing $H_\theta$ instead of $H$ in order to emphasize the value of the $\theta$ parameter, the commutation relations \eqref{Q5H} as well as the fact that additional commutators with $Q_5$ vanish, implies 
\es{VHV}{
	{\cal V}_\alpha H_\theta  {\cal V}_{\alpha}^{-1} &=
	H_\theta + i \alpha [Q_5, H_\theta] 
	- \frac{\alpha^2}{2} [Q_5, [Q_5, H_\theta] ] \\
	&=  H_{\theta -2 \alpha} \,.
}
Thus,  the Hamiltonians with different values of $\theta$ are equivalent up to the unitary transformation by $ {\cal V}_\alpha$, and therefore the energy spectrum is independent of $\theta$.  The Schwinger anomaly can therefore be used to set $\theta = 0$.

\section{More details on the lattice model}\label{sec:HamiltonianAppendix}

\subsection{Gauge symmetry and Gauss law}

Classically, the Hamiltonian \eqref{eq:HLattice} is invariant under gauge transformations that act as 
\es{GaugeOpsAppendix}{
	\chi_n \to V_n  \chi_n \,, \qquad U_n \to V_n U_n V_{n+1}^\dag \,,\\
	\chi_n^\dagger \to V_n^\dag  \chi_n^\dag \,, \qquad U_n^\dag \to V_{n+1} U_n^\dag V_n \,,
}
where the $V_n = e^{i v_n}$ are phases denoted collectively by $V$.  Quantum mechanically, such an action is represented by conjugation by a unitary operator
\es{UnitaryGauge}{
	{\cal U}_{V} = \exp \left[i \sum_{n=0}^{N-1} \left( -v_n Q_n + (v_n- v_{n+1} ) L_n  \right)  \right] \,,
}
where $Q_n$ are the charge operators introduced in \eqref{Gauss}.  Indeed, the fermion anti-commutation relations in \eqref{CommutLattice} imply $[Q_n, \chi_m] = -\delta_{nm} \chi_n$ and $[Q_n, \chi_m^\dagger] = \delta_{nm} \chi_n^\dagger$, which, together with the commutators between $L_n$ and $U_m$ and $U_m^\dagger$ imply that
\es{ConjugWithV}{
	{\cal U}_{V} \chi_n  {\cal U}_{V}^{-1} = V_n \chi_n \,, \qquad
	{\cal U}_{V} U_n  {\cal U}_{V}^{-1} = V_n U_n V_{n+1}^\dag \,, \\
	{\cal U}_{V} \chi_n^\dag  {\cal U}_{V}^{-1} = V_n^\dag \chi_n^\dag   \,, \qquad
	{\cal U}_{V} U_n^\dag  {\cal U}_{V}^{-1} = V_{n+1} U_n^\dag V_n^\dag \,,
}
reproducing \eqref{GaugeOpsAppendix}.  

Physical states must obey ${\cal U}_V = 1$ for any $V$.  If the $v_n$ are all infinitesimal, this means that the exponent in \eqref{UnitaryGauge} must vanish identically on gauge invariant states.  This is equivalent to 
\es{Vanishing}{
	\sum_{n=0}^{N-1} v_n (L_n - L_{n-1} - Q_n)= 0
} 
for any $v_n$, which in turn implies the Gauss law \eqref{Gauss}. As in the continuum case, the Gauss law constraint does not imply ${\cal U}_V = 1$, and we also need to impose invariance under large gauge transformations.  In this case, we can take, for instance, $V_{n, \text{large}} = e^{i v_{n, \text{large}}}$ with $v_{n, \text{large}} = 2 \pi n / N$.  We have
\es{UVLarge}{
	{\cal U}_{V_\text{large}} =  \exp \left[-2 \pi i \frac 1N \sum_{n=0}^{N-1} \left( n Q_n +  L_n  \right)  \right] \,.
}
Using \eqref{Gauss}, we find
\es{UVLargeAgain}{
	{\cal U}_{V_\text{large}} =  e^{- 2 \pi i L_{N-1}} \,.
}
Thus, invariance under large gauge transformations also implies that $L_{N-1} \in \Z$ when acting on any physical state.  Here, the ($N-1$)st site was chosen arbitrarily, so we must have that 
\es{LargeGaugeL}{
	e^{2 \pi i L_n} = 1 
}
when acting on gauge invariant states, for any given $n$.  In other words, the operators $L_n$ should have integer eigenvalues.

A related property of the model \eqref{eq:HLattice} is that the theories in which $\theta$ differs by a multiple of $2 \pi$ are equivalent.  Physically, this is because $\theta / 2 \pi$ should be interpreted as the fractional part of the effective electric field (see also~\eqref{Eeff})
\es{Leff}{ 
	L_{n, \text{eff}} = L_n + \frac{\theta}{2 \pi} \,.
}
As in the continuum model the fact that this is the effective electric field can be seen from the first term in the Hamiltonian \eqref{eq:HLattice}.  A more mathematical derivation of the fact that $\theta$ has period $2 \pi$ is as follows.  Consider the holonomy ${\cal H} \equiv U_0 U_1 \cdots U_{N-1}$, which is a unitary operator.  Conjugation by it shifts all $L_n$ by $-1$, which implies that
\es{Holtheta}{
	{\cal H} H_{\theta}  {\cal H}^{-1} = H_{\theta - 2 \pi} \,.
}
Thus, the Hamiltonians with $\theta$ parameters that differ by multiples of $2 \pi$ are unitarily equivalent.

\subsection{Eliminating the gauge field}

If we were working with OBC, Eq.~\eqref{Gauss} would allow us to completely integrate out the gauge field. With PBC, however, the best we can do is to eliminate all gauge degrees of freedom except for the holonomy variable and its conjugate, the average electric field.  

Indeed, we can solve \eqref{Gauss} by introducing the average electric field
\begin{equation}
	{\cal E} \equiv \frac{1}{N} \sum_{n=0}^{N-1} L_n \,,
\end{equation}
and expressing all the link variables in terms of $\mathcal{E}$ and the site charges $Q$:
\es{GotLn}{
	L_n &= {\cal E} + \frac 1N \sum_{m=1}^{N} (m - N \theta_{m>n}) Q_{m} \,, \\
	&\text{where} \quad \theta_{m>n} \equiv \begin{cases}
		1 & \text{if $m>n$} \\
		0 & \text{otherwise} \,.
	\end{cases}
}
At the same time, we can perform a unitary transformation on the Hilbert space, using an appropriate ${\cal U}_{V}$ that sets $U_n = U$ for all $n$.  The variable $U^N$ is canonically conjugate to ${\cal E}$; the two obey the commutation relation $[{\cal E}, U^N] = U^N$. The Hamiltonian then becomes
\es{HLattice3}{
	H &= \frac{e^2 N a}{2} \left({\cal E} + \frac{\theta}{2\pi}\right)^2 + m_\text{lat}  \sum_n (-1)^{n}\chi_n^\dag \chi_n\\
	&{}- \frac{i}{2a} \sum_{n=0}^{N-1} \left[ \chi_n^\dag U \chi_{n+1}-\chi_{n+1}^\dag U^\dag \chi_{n}\right]   \\
	&{}- \frac{e^2 a}{4N}  \sum_k \sum_{k'} \abs{k - k'}  \left(N-\abs{k - k'} \right)  Q_{k} Q_{k'} \,.
} 
This is a simplified model that is equivalent to the original Hamiltonian \eqref{eq:HLattice}.

Instead of \eqref{Basis}, one can now consider a basis of states labeled by the site occupation numbers $n_m \in \{0,1\}$ %
and average electric field $\mathcal{E} \in \frac{1}{N}\mathbb{Z}$:
\begin{equation}\label{eq:states}
	\ket{\mathcal{E};n_0 n_1 \cdots n_{N-1}} \equiv \left( \prod_{m=0}^{N-1} \left( \chi_m^\dagger\right)^{n_m} \right) U^{N {\cal E}} \ket{\text{vac}} \,.
\end{equation}
All that remains of \eqref{Gauss} is the global constraint, implied by consistency of PBC:
\begin{equation}
	\sum_{m=0}^{N-1} n_m = \frac{N}{2} \,.
\end{equation}
In addition, one should also impose invariance under the large gauge transformations as in \eqref{LargeGaugeL}.  For $n=0$, this implies
\es{InvLargeCalE}{
	{\cal E} + \frac 1N \sum_{m=0}^{N-1} m Q_m \in \Z \,.
}
Thus, for gauge-invariant states, ${\cal E}$ cannot be any multiple of $1/N$, but instead only those multiples of $1/N$ that obey \eqref{InvLargeCalE}.

\subsection{Comments on the charge $q$ model}

As mentioned in the Introduction, if we made the redefinitions
\es{Redefs}{ 
	U_n &= (U_n')^q \,, \qquad L_n = L_n' / q \,, \qquad e = e' q\,, \\
	\theta &= \theta' / q\,, \qquad m_\text{lat} = m_\text{lat}' \,, \qquad
	\chi_n = \chi_n' \,, 
}
then, in terms of the primed variables, we obtain a model in which the fermions have charge $q$. The Hamiltonian of the charge-$q$ model is 
\es{eq:HLatticeChargeq}{
	H' &= \frac{e'^2 a}{2} \sum_{n=0}^{N-1} \left( L_n' + \frac{\theta'}{2 \pi} \right) ^2 + m_{\rm lat}' \sum_{n=0}^{N-1} (-1)^{n} \chi^{\prime \dag}_n \chi_n'  \\
	&- \frac{i}{2a} \sum_{n=0}^{N-1} \left[ \chi_n^{\prime \dag} U_n'^q \chi_{n+1}- \chi_{n+1}^{\prime \dagger} (U_n^{\prime \dagger})^q \chi_n' \right]  \,,
}
and the Gauss law is
\es{GaussChargeq}{
	L_{n}'-L_{n-1}' = Q_n' \,, \qquad Q_n' \equiv q \left[ \chi_n^{\prime \dag} \chi_n' -\frac{1-(-1)^n}{2} \right] \, .
}

Classically, the gauge transformations that leave $H'$ invariant are
\es{GaugeChargeq}{
	\chi_n' \to (V_n')^q  \chi_n' \,, \qquad U_n' \to V_n' U_n' V_{n+1}^{\prime \dag} \,,\\
	\chi_n^{\prime \dagger} \to (V_n^{\prime \dag})^q  \chi_n^{\prime \dag} \,, \qquad U_n^{\prime \dag} \to V_{n+1}' U_n^{\prime \dag} V_n' \,,
}
and, just as in the unit charge case, they are implemented by the unitary operator 
\es{UnitaryGaugeChargeq}{
	{\cal U}'_{V'} = \exp \left[i \sum_{n=0}^{N-1} \left( -v_n' Q_n' + (v_n'- v_{n+1}' ) L_n'  \right)  \right] \,,
}
where $V_n' = e^{i v_n'}$.  The requirement that ${\cal U}'_{V'}  = 1$ on the gauge-invariant states implies the Gauss law \eqref{GaussChargeq}, as well as the fact that $e^{2 \pi i L_n'} = 1$ for any given $n$.  Thus, the electric fields $L_n'$ must be integer. 

An interesting feature of the charge-$q$ model is that it has a lattice $\Z_q$ one-form symmetry generated by the family of unitary topological operators
\es{calW}{
	{\cal W}_n' \equiv e^{ 2 \pi i L_n' / q} \,.
}
The symmetry is $\Z_q$ because $({\cal W}_n')^q = e^{2 \pi i L_n'} = 1$ on gauge-invariant states.   
The operators ${\cal W}_n'$ essentially measure the electric field mod $q$.  They are ``topological'' (i.e.~independent of $n$) because the fermions have charge $q$, and so the electric field mod $q$ is the same on all sites.  Thus ${\cal W}_n' = {\cal W}_m'$ for any $n$, $m$, when acting on physical states. It is also not hard to see that for any given $n$, we have
\es{calWH}{
	{\cal W}_n' H' ({\cal W}_n')^{-1} = H'  \quad
	\Longleftrightarrow \quad [ {\cal W}_n' , H' ] = 0 \,, 
}
implying that ${\cal W}_n'$ generate a symmetry.  The last equation follows from the conjugation relations 
\es{ConjugByW}{
	{\cal W}_n' \chi_m' ({\cal W}_n')^{-1} &= \chi_m' \,, \\
	{\cal W}_n' L_m' ({\cal W}_n')^{-1} &= L_m' \,, \\
	{\cal W}_n' U_m' ({\cal W}_n')^{-1} &= e^{ 2 \pi i \delta_{nm} / q} U_m' \,, 
}
as well as the hermitian conjugates of these expressions, and the expression for $H'$ in \eqref{eq:HLatticeChargeq}.

Because ${\cal W}_n'$ commute with the Hamiltonian, they are simultaneously diagonalizable, so the spectrum of the Hamiltonian splits into sectors where ${\cal W}_n' = e^{2 \pi i r / q}$, with $r = 0, 1, \ldots, q-1$.  The sector of a given $r$ has been referred to as a ``universe'' in the context of continuum theories \cite{Hellerman:2006zs,Komargodski:2020mxz} (see also \cite{Misumi:2019dwq,Cherman:2021nox}).  In terms of the original model before the rescaling \eqref{Redefs}, the $r$th universe corresponds to the theory with $\theta = (2 \pi r + \theta') / q$.  While in the unit charge model, the values of $\theta = (2 \pi r + \theta') / q$, with $r = 0, 1, \ldots, q-1$ correspond to different theories with distinct Hamiltonians, in the charge $q$ model all these Hamiltonians are combined into a block-diagonal Hamiltonian of a single theory.

The discussion so far applies to the charge-$q$ model with any $m_\text{lat}'$.  When $m_\text{lat}' = -q^2 e^2 a / 8$, we can again define unitary operators ${\cal V}'$ that implement the translation by one site.  As in the unit charge case, one can show that
\es{VHVChargeq}{
	{\cal V}' H'_{\theta'} {\cal V}^{\prime -1} = H'_{\theta' + q \pi} \,.
}

Note that as in the unit charge model, conjugation by the holonomy ${\cal H}' = U_0' U_1' \cdots U_{N-1}'$ shifts all $L_n'$ by $-1$.  From the definition of the Hamiltonian \eqref{eq:HLatticeChargeq} it follows that
\es{Holthetaprime}{
	{\cal H}' H'_{\theta'}  {\cal H}^{\prime -1} = H'_{\theta' - 2 \pi} \,.
}
Thus, when $q$ is even, the shift in $\theta'$ in \eqref{VHVChargeq} can be undone by conjugation with $({\cal H}')^{q/2}$, so in this case the unitary transformation implemented by  ${\cal V}' ({\cal H}')^{q/2}$, namely translation by one site combined with a shift of the electric fields by $-q/2$ is a $\Z_2$ symmetry of the Hamiltonian.  This is a $\Z_2$ subgroup of the $\Z_q$ chiral symmetry of the charge-$q$ Schwinger model \cite{Armoni:2018bga,Misumi:2019dwq,Komargodski:2020mxz}.

\subsection{Comparison with \cite{Berruto:1997jv}}

Let us now go back to the $q=1$ model and explore the connection between this model and that presented by Berruto {\em et al.} in \cite{Berruto:1997jv}.  Ref.~\cite{Berruto:1997jv} considered the lattice Hamiltonian (written after redefining $e_L \to e$, $x \to n$, $\psi_n \to \tilde \chi_n$, $e^{i A_n} \to \tilde U^\dagger$, $E_n \to -\tilde L_n$, and $t = - 2 a w$):
\es{HamiltBerruto}{
	H_B &= \frac{e^2 a}{2} \sum_n \tilde L_n^2 \\
	&{}- iw \sum_n \left( \tilde \chi_n^\dagger \tilde U_n \tilde \chi_{n+1} 
	-  \tilde \chi_{n+1}^\dagger \tilde U_n^\dagger \tilde \chi_n \right) \,,
}
which is the same as our Hamiltonian \eqref{eq:HLattice} with $m_\text{lat}= \theta = 0$ and all operators replaced by their tilded counterparts.  However,  Ref.~\cite{Berruto:1997jv} considered a non-staggered Gauss law constraint, which after the redefinitions written above is
\es{GaussBerruto}{
	\tilde L_n - \tilde L_{n-1}  =  \tilde \chi_n^\dagger \tilde \chi_n - \frac 12  \,.
}
The commutation relations of the various operators are the same as those in \eqref{CommutLattice} after placing tildes over all operators.  With this non-staggered constraint, Ref.~\cite{Berruto:1997jv} noticed that the Hamiltonian is invariant under a discrete one-site translation symmetry
\es{TransSym}{
	\tilde U_n &\to U_{n+1} \,, \qquad  \tilde L_n \to L_{n+1} \,, \\
	\tilde \chi_n &\to \chi_{n+1}  \,, \qquad \tilde \chi_n^\dag \to \tilde \chi_{n+1}^\dag \,.
} 

To see why this model is the same as \eqref{eq:HLattice} with our Gauss law constraint \eqref{Gauss} and $m_\text{lat} = - e^2 a / 8$, we can define
\es{opRedef}{
	L_n &= \tilde L_n  + \frac{(-1)^n}{4} - \frac{ \theta}{2 \pi} \,,\\
	U_n &= \tilde U_n \,, \qquad \chi_n = \tilde \chi_n  \,.
}
Then, the Hamiltonian \eqref{HamiltBerruto} becomes
\es{OurHBer}{
	H_B &= \frac{e^2 a}{2}  \sum_n  \left( L_n + \frac{\theta}{2 \pi} \right)^2  - \frac{e^2 a}{2}  \sum_n L_n \frac{(-1)^n}{2} \\
	&{}+ \frac{e^2 a N}{32} -i w \sum_n \left( \chi_n^\dagger U_n \chi_{n+1} -\chi_{n+1}^\dagger U_n^\dagger \chi_n  \right) \,,
}
and the Gauss law becomes identical to \eqref{Gauss}. Using the Gauss law \eqref{Gauss}, the second term in \eqref{OurHBer} can be written in terms of $\chi_n^\dagger \chi_n$, giving the final result
\es{OurHBerAgain}{
	H_B &= \frac{e^2 a}{2}  \sum_n  \left( L_n + \frac{\theta}{2 \pi} \right)^2  - \frac{e^2 a }{8} \sum_n (-1)^n \chi_n^\dagger \chi_n \\
	&{} -i w \sum_n \left( \chi_n^\dagger U_n \chi_{n+1} -\chi_{n+1}^\dagger U_n^\dagger \chi_n  \right)- \frac{e^2 a N}{32} \,.
}
Up to the constant shift by $-e^2 a N/32$, this is precisely our Hamiltonian \eqref{eq:HLattice} at the special point  $m_\text{lat} = - e^2 a / 8$.

So far, we left the parameter $\theta$ arbitrary, but the value of $\theta$ is determined by requiring that the states we are interested in have $e^{2 \pi i L_n} = 1$.

Ref.~\cite{Berruto:1997jv} considered two states  written in Eqs.~(2.4) and (2.5) of that paper, which in our notation are
\es{PsiChiDef}{
	\ket{\psi} &= \left( \prod_{\text{$n$ even}} \chi_n^\dagger \right) \left( \prod_n U_n^{\frac{(-1)^n}{4}} \right) 
	\ket{\text{vac}}_B \,, \\
	\ket{\chi} &= \left( \prod_{\text{$n$ odd}} \chi_n^\dagger \right) \left( \prod_n U_n^{\frac{-(-1)^n}{4}} \right) 
	\ket{\text{vac}}_B \,,
}
where $\ket{\text{vac}}_B$ is the vacuum considered in \cite{Berruto:1997jv}.  A subtlety is that this vacuum is different from the vacuum $\ket{\text{vac}}$ we consider, because the former is annihilated by all $\tilde L_n$, while the latter is annihilated by all $L_n$.  (Both are annihilated by all $\tilde \chi_n = \chi_n$.)  Given \eqref{opRedef}, it follows that 
\es{vacBExplicit}{
	\ket{\text{vac}}_B = \left( \prod_n U_n^{\frac{(-1)^n}{4} - \frac{\theta}{2 \pi}}  \right) \ket{\text{vac}} \,.
}
Combining \eqref{PsiChiDef} with \eqref{vacBExplicit}, we obtain
\es{PsiChiAgain}{
	\ket{\psi} &= \left( \prod_{\text{$n$ even}} \chi_n^\dagger \right) 
	\left( \prod_n U_n^{\frac{(-1)^n}{2}  - \frac{\theta}{2 \pi}} \right) 
	\ket{\text{vac}} \,, \\
	\ket{\chi} &= \left( \prod_{\text{$n$ odd}} \chi_n^\dagger \right) \left( \prod_n U_n^{ - \frac{\theta}{2 \pi}} \right) 
	\ket{\text{vac}} \,.
}
For these states to be gauge invariant, we should choose $\theta = \pi$ for $\ket{\psi}$ and $\theta = 0$ for $\ket{\chi}$:
\es{PsiChiAgain2}{
	\ket{\psi} &= \left( \prod_{\text{$n$ even}} \chi_n^\dagger \right) 
	\left( \prod_{\text{$n$ odd}}  U_n^{-1} \right) 
	\ket{\text{vac}} \,, \quad \theta = \pi \\
	\ket{\chi} &= \left( \prod_{\text{$n$ odd}} \chi_n^\dagger \right) 
	\ket{\text{vac}} \,, \quad \theta = 0 \,.
}

\onecolumngrid
\vspace{1in}
\twocolumngrid

\bibliographystyle{ssg}
\bibliography{Schwinger_draft,Schwinger_draftNotes}

\end{document}